\documentclass[12pt]{iopart}
\usepackage{iopams,epsf,graphicx}  
\begin{document}

\title{Monte Carlo study of the Pure and Dilute Baxter-Wu model}

\author{Nir Schreiber
 \ and Joan Adler 
}

\address{Physics Department, Technion, Israel Institute of Technology, 
Haifa, Israel, 32000  }


\begin{abstract}
We studied the pure and  dilute Baxter-Wu (BW) models 
using the Wang-Landau (WL)
sampling method to calculate the Density-Of-States (DOS).
We first used the exact result for the DOS of the Ising model to test our code. Then
we calculated the DOS of the dilute Ising model to obtain a phase diagram,
in good agreement with previous studies.
We calculated the energy distribution, together  
with its first, second and fourth moments,
to give the specific heat and the 
energy fourth order cumulant, better known as the Binder parameter, for the pure BW model. 
For small samples, the energy distribution displayed a doubly peaked shape, and
 finite size scaling analysis 
 showed the expected reciprocal scaling of the positions of the peaks with $L$.
The energy distribution yielded the expected BW $\alpha=2/3$ critical exponent for the specific heat. The Binder parameter minimum appeared to scale with lattice size $L$ 
with an exponent $\theta_B$ equal to the specific heat exponent. Its location (temperature) 
showed a large correction-to-scaling term $\theta_1=0.248\pm0.025$. For the dilute BW model we found a clear crossover to a single peak 
in the energy distribution even for small sizes  and 
 the expected
$\alpha=0$ was recovered. 

\end{abstract}


\submitto{\JPA}

\maketitle

\section{Introduction}
The two-dimensional 
Ising model has received such widespread attention as the paradigm
system for phase transitions, that one sometimes says that a certain
system is the ``Ising model'' of a class of problems.
While it is clearly special,  it is not the only two dimensional model
of phase transitions with an exact expression for its free energy. 
Its critical behavior is, in fact, rather atypical 
relative to many other
two-dimensional systems
and even to the three-dimensional Ising model, especially
in the specific heat,
where its critical exponent, $\alpha$, is zero.  The nature of the
 corrections-to-scaling in the spin 1/2 Ising model
is also very different to that of many other
 interesting systems.

\begin{figure}[h]
\begin{center}
\includegraphics[width=5.00cm]{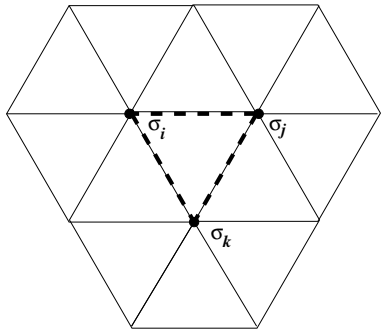}
\end{center}
\caption{\label{fig:BW}The energy of a given configuration is the sum of all interacting 
triangles formed by nearest neighbor spins}
\end{figure}

Another spin system, now known as the
 Baxter Wu  (BW) model, 
  was solved  by R.J. Baxter
 and F.Y. Wu~\cite{BW1,BW2}.
Spins $\sigma_i=\pm1$,
are situated on the triangular lattice
and interact via a three spin interaction, 
\begin{equation}
{\cal{H}}=-J\sum_{i,j,k}\sigma_i\sigma_j\sigma_k,
\end{equation}
where $i,j$ and $k$ are the vertices of a
 triangle as shown in Fig.~\ref{fig:BW}. $J>0$ is the ferromagnetic
coupling between nearest neighbor spins.
The BW model exhibits a second order phase
transition with its 
critical temperature ($T_c$) given by 
$2J/kT_c=\ln(1+\sqrt 2)=2.26918...$, 
(the same numerical value  as for 
  the Ising model on the square lattice). The specific heat critical exponent is equal to the correlation length
 exponent, $\alpha=\nu=2/3$. 
Series-expansion results~\cite{BaxterSykesWatts}, gave the conjectured 
magnetization exponent of $\beta=1/12$ and
a susceptibility  exponent of $\gamma\approx1.17$~\cite{Griffiths}. 
The latter confirmed the prediction of $\gamma=7/6$ from the well known
 scaling relation
$\alpha+2\beta+\gamma=2$~\cite{Essam,Rushbrooke}.
 Real Space Renormalization Group methods 
have also been used~\cite{Braathen,Imbro,Nijs} to study the pure model,
and the critical eigenvalues obtained gave critical exponents 
consistent with series-expansion and exact results.
 An exact form for  BW corrections-to-scaling  was found 
by Joyce~\cite{Joyce}
 who conjectured
that the
 spontaneous magnetization varied as 
$M=t^\beta\left(f_0(t) +t^{2/3}f_1(t)+\cdots\right)$
with analytic functions $f_0, \, f_1$ of the distance
$t=(T-T_c)/T_c$. Adler and Stauffer confirmed this with
series and Metropolis Monte Carlo estimates~\cite{adst}. 

Dilute Ising models are also somewhat famous, but for rather 
different reasons, as they have been the source of a great deal of controversy.
Presumably because of the anomalous specific heat structure in the 
pure case, numerical work in the dilute regime, especially 
near the pure limit is painful, and although 
a majority of authors (see e.g. Roder $et \ al$,~\cite{Roder}) have
claimed that the controversy is resolved in favor 
of SSL theory~\cite{Shalaev,Shankar,Ludwig}, 
more study is useful.

The annealed dilute BW model was studied by Kinzel, Domany and Aharony\cite{aharony}
who showed from this exploration 
that its dominant critical behavior is in the 
universality class of the four state Potts model, although the Potts 
model has logarithmic correction terms for this case. Domany and Riedel~\cite{Domany}, argued the same for 
the pure BW model by means of symmetries of the Landau-Ginzburg-Wilson Hamiltonian.
 (Note that there 
are first order fixed points in the neighborhood of these models).
The quenched dilute
 BW model was studied by Landau and Novotny~\cite{Novotny}, who found
 a substantial change in the critical behavior
of the specific heat~\cite{Harris} for an impurity concentration of
 $1-x=0.1$. They also 
conjectured that the zero temperature threshold
concentration above which no long-range order could be seen was
 about $x_c\simeq0.71$. (See also results from a cluster-algorithm study  
of this system~\cite{Evertz}).
More recent calculations~\cite{fried,adler} showed that the value of $x_c$ is 
 even higher ($x_c\simeq0.755$), and is bounded by $x_c^{\rm low}=0.710\pm0.001$ and
 $x_c^{\rm high}=0.784\pm0.004$. This is substantially above the value for Ising models where $x_c$ is simply
 the percolation threshold of the corresponding lattice, which is rarely above 0.5 .

Recently Wang and Landau~\cite{WL1,WL2} proposed a very efficient
algorithm for calculating the density-of-states (DOS), (i.e. the degeneracy of any level in energy space),
$g(E)$,
for Ising models and some related systems.
To explore
  the issues of both pure and dilute BW models further,
 and to see how its different
particulars of large $\alpha$ and corrections to scaling emerge from
the calculation of the DOS,
we have chosen to apply the WL algorithm to both the
pure and quenched dilute BW
 models, and to study the behavior of the energy distribution and related moments~\cite{Challa,histogram2}
 using the simulated DOS.
 The DOS of the pure and dilute Ising models was studied for comparison purposes.

In the next section we discuss the WL algorithm.
 In section~\ref{sec:Ising} we
 present a comparison of an exact calculation of the DOS for the
Ising model~\cite{Beale}
with simulations using WL and give some results for the dilute Ising case.
 In section~\ref{sec:BW_pure}
 we give in detail our results for the pure BW model,
and in section ~\ref{sec:BW_dilute}
the results for the dilute BW model are presented.
Finally we discuss the implications of our results in
section~\ref{sec:discussion}.

\section{ The simulation method}

Conventional Monte-Carlo (MC) methods~\cite{Metropolis,Swendsen,Wolff}
generate the canonical energy distribution at a given
temperature $T_0$. It is usually narrowly peaked around this temperature.
The need to perform
multiple simulations in order to obtain thermodynamics
in a large range of temperatures requires a large
computational effort.
Other methods based on histogram accumulation
~\cite{histogram2,histogram1}
approximate the distribution by the energy histogram at $T_0$.
This distribution can then be reweighted
to give statistics at another temperature. The reweighted
 distributions, however, are also
restricted to a very narrow range of temperatures and suffer
 from large statistical errors in their tails for
temperatures far from $T_0$.
The broad histogram method~\cite{deOliveira}
calculates the DOS through the consideration of the average number of visits to
any two adjacent energy levels. Lee~\cite{Lee} offered the entropic sampling 
method using the observation that if the transition probability between any
two energy levels is proportional to the ratio between the DOS
of these levels, then a crude estimate to the DOS can be given 
when sampling at infinite temperature.

Wang and Landau improved Lee's method by introducing a modification factor
which together with generating a ``flat'' histogram 
(we have used the condition $|H(E)-\langle H\rangle|/\langle H\rangle \le0.05$ for any $E$),
carefully controls the updating of the DOS.  
By dividing the energy space into
 different segments, and
performing an independent random walk in each segment,
one can generate very accurately, in a reasonable amount of CPU time,
the DOS of the whole energy space,
thus obtaining the canonical distribution at any desired temperature.
\section{Ising model results}{\label{sec:Ising}}
\subsection{The pure Ising model}
We began by validating the accuracy of our implementation of
 the WL algorithm against exact
 results for the Ising model on the square lattice with no impurities.
A detailed comparison was made for the case of $L=32$.
The partition function for the Ising model  on a lattice of length
$L$ can be written as a low
temperature expansion
\begin{equation}
{\cal{Z}}_N=e^{2KN}\sum_{\ell}g_{\ell}x^{2\ell},
\label{Z_lowT}
\end{equation}
where $N=L\times{L}$ is the number of spins, $K=J/kT$ is
the reduced inverse temperature,
and $x=e^{-2K}$ is the low temperature variable. Each energy level can be labeled (relative to the ground state energy $-2JN$) by $E_{\ell}=4J\ell$ $(\ell=0,2,3,...N-2,N)$,
so that $g_{\ell}$ is its corresponding DOS.
Beale~\cite{Beale} used an  extension of Onsager's solution~\cite{Onsager}, to give
 the exact expression
for the partition function on a finite lattice~\cite{Kaufman},
and extracted the DOS coefficients from the
expansion~(\ref{Z_lowT}). When we plotted our results 
on top of Beale's expression for the case of $L=32$ we saw no
deviations between the exact  and the simulated data within the resolution
of the figure. The relative error between the exact and simulated data was
also plotted and was found to be three orders of magnitude smaller than 
the calculated DOS and two orders of magnitude larger from
the systematic error due to the choice of the final modification factor
$f_{\rm final}=0.001$. This showed that the choice of this 
quite large $f_{\rm final}$ was sufficient, so that only a relatively small
number of iterations was required for all the simulations performed
throughout this work.

Further results from the pure Ising simulations will be introduced for comparison purposes in section~\ref{sec:BW_pure}.

\subsection{The dilute Ising model}
\begin{figure}[h]
\begin{center}
\includegraphics[height=9.0cm,angle=-90]{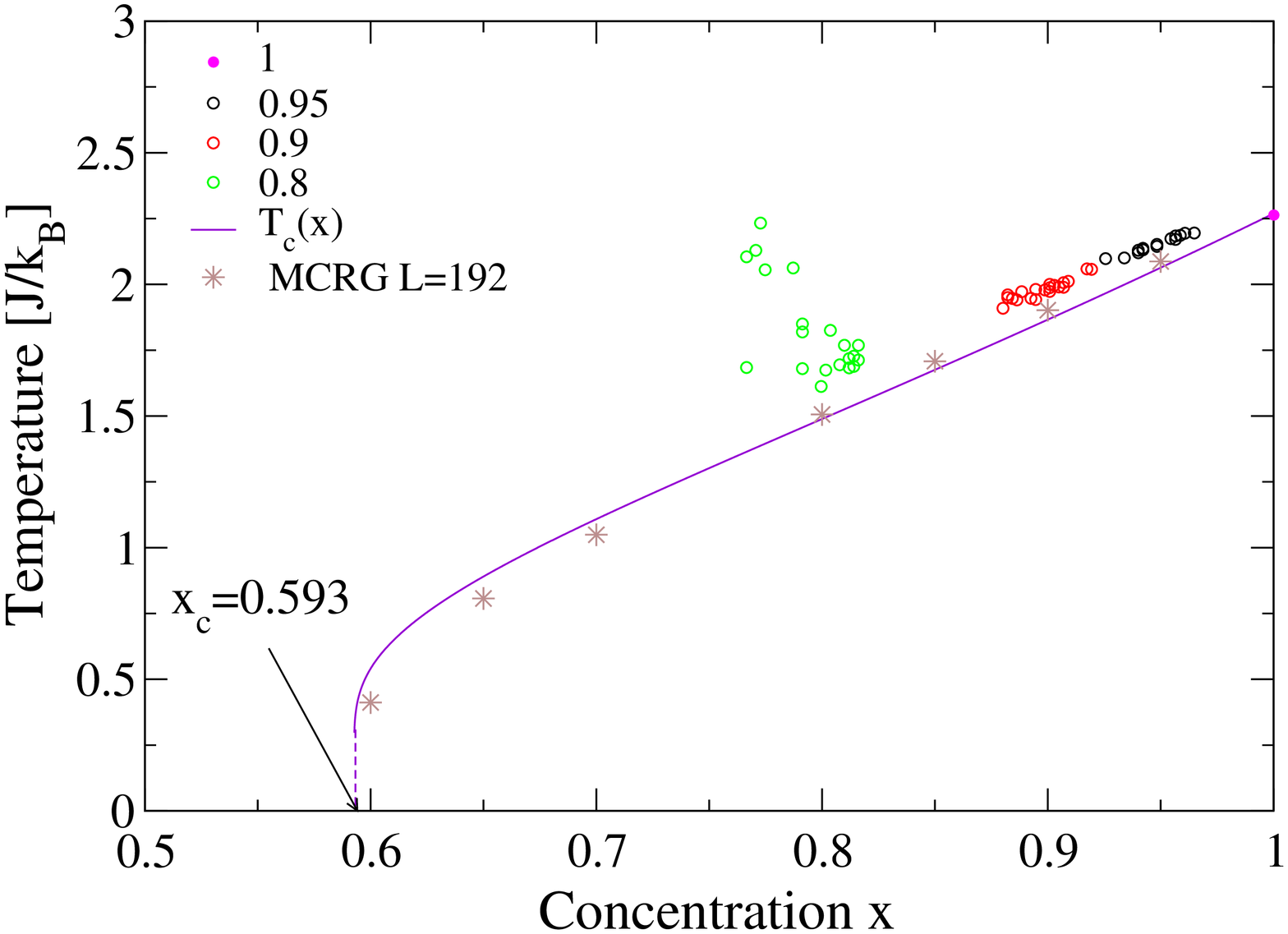}
\caption{Critical line $T_c(x)$ in the $T$-$x$ plane of the dilute Ising model. Small energy fluctuations for $x\leq0.8$ make it hard to reliably determine $T_{C_{\rm{max}}}$.
The asterisks represent results from MC Renormalization Group calculations~\cite{deSouza} and the solid line
is the prediction $T_c(x)=\{\tanh^{-1}[e^{-1.45(x-x_c)}]\}^{-1}$, converging to the value of $x_c=p_c=0.593$~\cite{Stauffer}.}
\label{fig:Idilute}
\includegraphics[height=9.0cm,angle=-90]{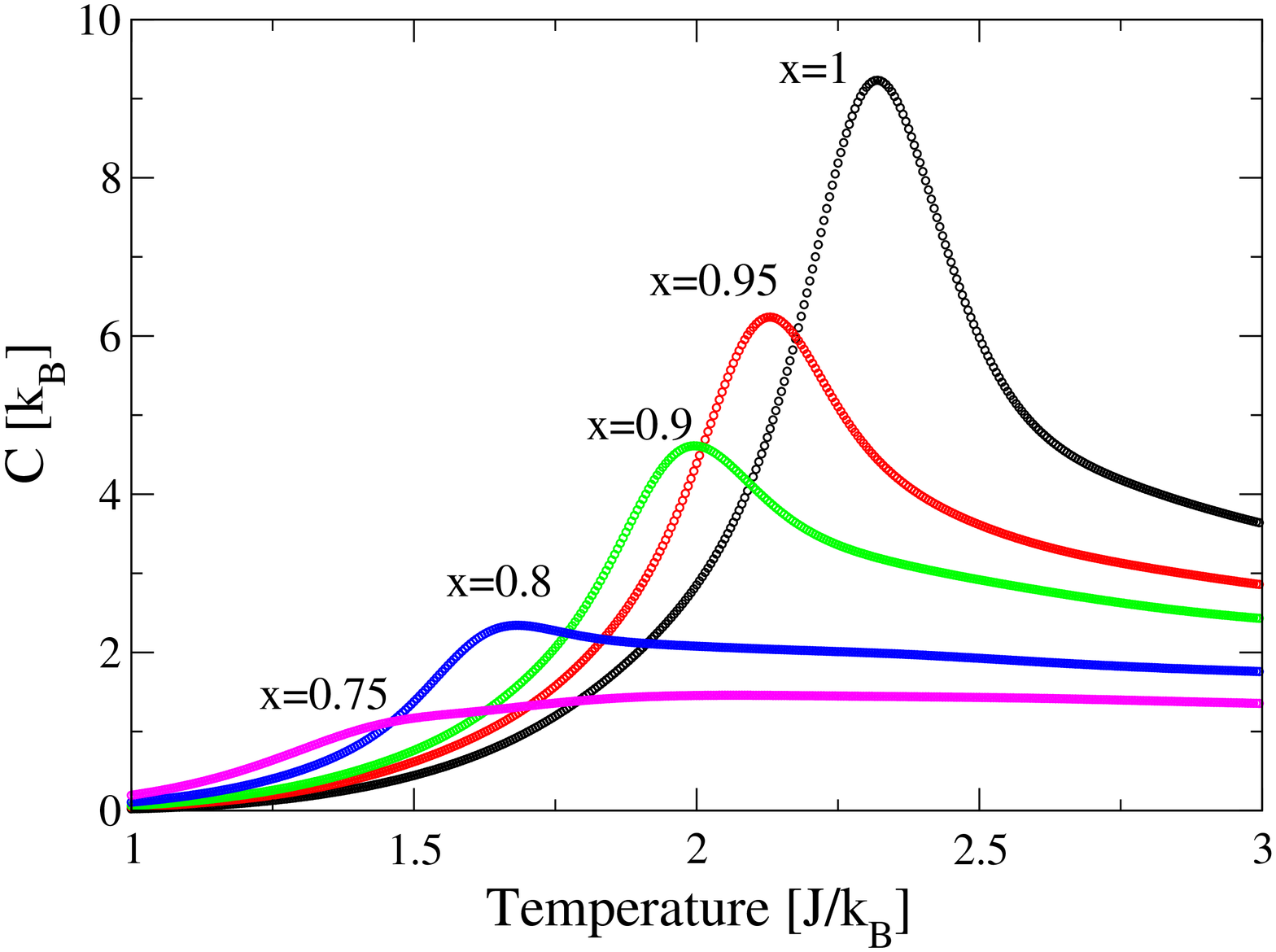}
\caption{The specific heat of the dilute Ising model for different concentrations on a $L=22$ lattice.
For $x=0.75$ there is no pronounced peak present}
\label{fig:CIdilute}
\end{center}
\end{figure}

We continued the validation process by studying the dilute Ising model.
at a lattice size of $L=22$.
The Hamiltonian for the dilute Ising model may be written as
\begin{equation}
{\cal{H}}=-J\sum_{i,j}\epsilon_i\epsilon_j\sigma_i\sigma_j,
\end{equation}
where the random disorder variables $\epsilon_i$ take the values 0 and 1,
such that their configurational average is equal to a dilution of  $0<x<1$.
We considered the position of the specific heat maxima, $T_{C_{\rm{max}}}$, for different nominal concentrations
 centered around the values $x=0.8,0,9$ and $x=0.95$, as
 indicated in Fig.~\ref{fig:Idilute}.
It is clearly seen that for large concentrations $(x\geq0.9)$ the
 circles tend to a continuously critical
line, slightly shifted from the solid line. The shift is a finite
size effect due to the use of a small sample.
For smaller concentrations there is a large dispersion of the circles
and the data are 
less reliable.
As shown in Fig.~\ref{fig:CIdilute}, the specific heat maximum becomes
broader with decreasing concentration and is hard to locate precisely.
 The reason for this is that when lowering the concentration 
isolated clusters which rarely interact
with each are formed, and hence
energy fluctuations become smaller. For a concentration of $x=0.75$ 
 these fluctuations are also nearly constant
and therefore no pronounced peak can be identified. It should be noted that 
presumably, when much larger samples would be used, a pronounced peak should be clearly
seen for concentration even lower than $x=0.75$ (see, for example~\cite{Selke}).
In the absence of analytic results, the location of our points
close to earlier estimates validates both our dilute code and our analysis methods.

\section{The pure Baxter-Wu model}{\label{sec:BW_pure}
We calculated the DOS for the BW model
using lattice sizes $L$ ranging from 6 to 120,
with periodic boundary conditions being imposed. For each lattice size the data was collected separately for each energy segment and then was combined to give the density of states for the entire energy landscape. We averaged over nine different runs for $L=30$, and saw that the fluctuations were
three orders of magnitude smaller then the measured quantity ($\ln g$), so that we neglected these fluctuations and for each
lattice size we executed a single run per segment only. By symmetry, for any state with negative energy, there exist a state with positive energy, so that it was sufficient to carry out the random walk only for non positive energies. (A similar argument holds for the Ising model). Plots of the internal energy, specific heat, free energy and entropy are given
in Fig.~\ref{fig:54results}.
\begin{figure}[h]
\begin{center}
\includegraphics[height=9.0cm,angle=-90]{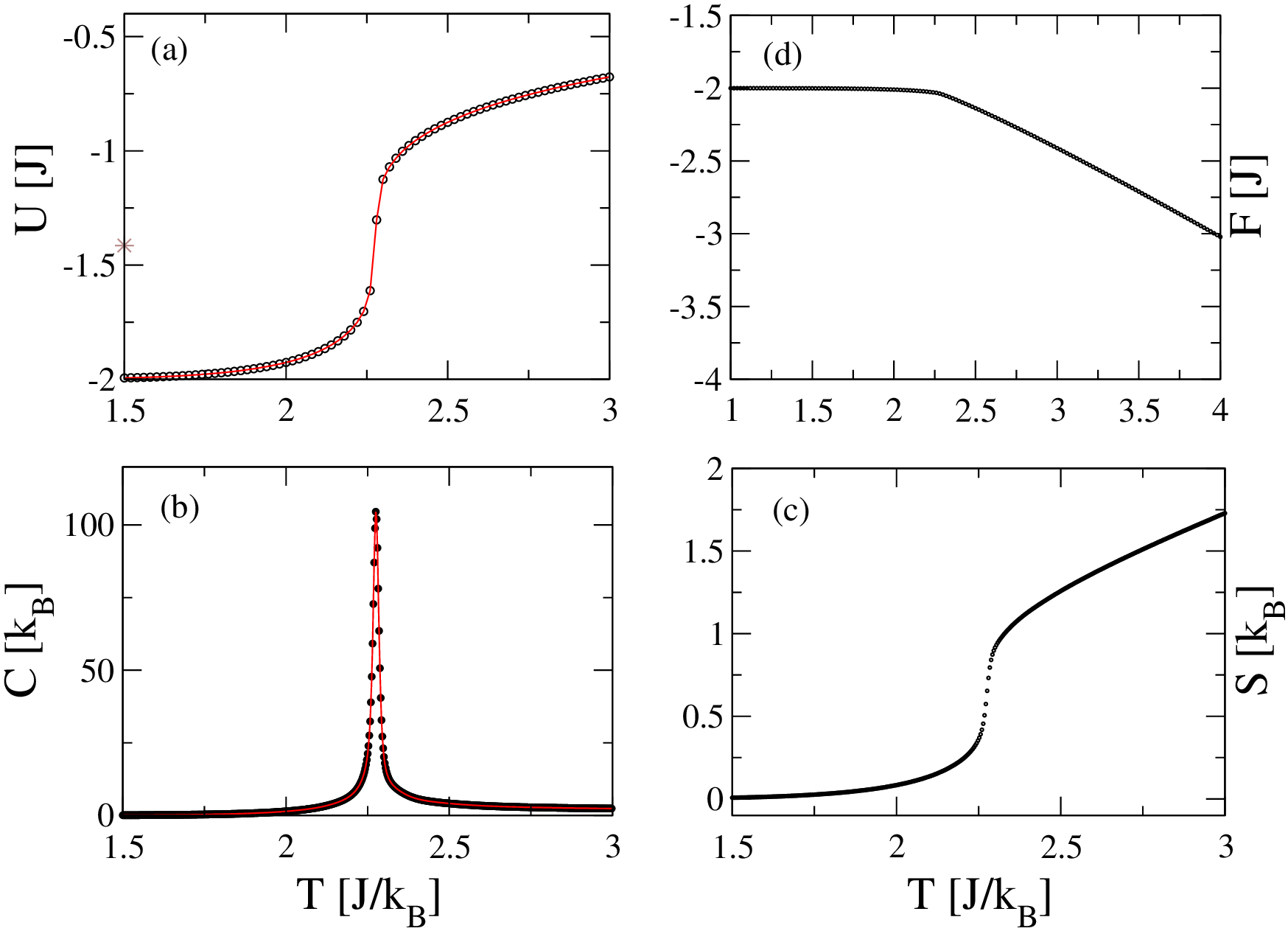}
\caption{Calculation of thermodynamic functions for the pure BW model on an $L=54$ lattice: (a) Internal Energy, (b) Specific Heat, (c) Entropy and (d) Free energy. The specific heat displays a very clear pronounced peak at the transition point.}
\label{fig:54results}
\end{center}
\end{figure}

Early simulations~\cite{Novotny} showed the formation and motion of domains around the ferromagnetic and ferrimagnetic ground
states, due to the special connectivity of the BW model, causing low frequency large energy fluctuations. These fluctuations made the impression that the system was in
a metastable state, thus indicating a first order transition.
In Fig~\ref{fig:H} we examined the energy distribution at $T_{C_{\rm{max}}}$ and found a doubly peaked curve
(see ref.~\cite{Kosterlitz}).
The system appears to fluctuate between these two peaks denoted by $E_-$, corresponding to an "ordered" state (more negative) energy, incorporating
small clusters, and $E_+$, corresponding to a "disordered" state energy incorporating large clustering.
A plot of the distribution for the Ising model both at $T_{C_{\rm{max}}}^{\rm{Ising}}$ and at $T_{C_{\rm{max}}}^{\rm{BW}}$
shows clearly sharp single peaks centered approximately at
the critical energy $U_c=-\sqrt2J$ (Fig.~\ref{fig:dos}).
This supports the uniqueness of the distributions in Fig~\ref{fig:H}.
\begin{figure}[h]
\begin{center}
\centerline{\includegraphics[height=9.00cm,angle=-90]{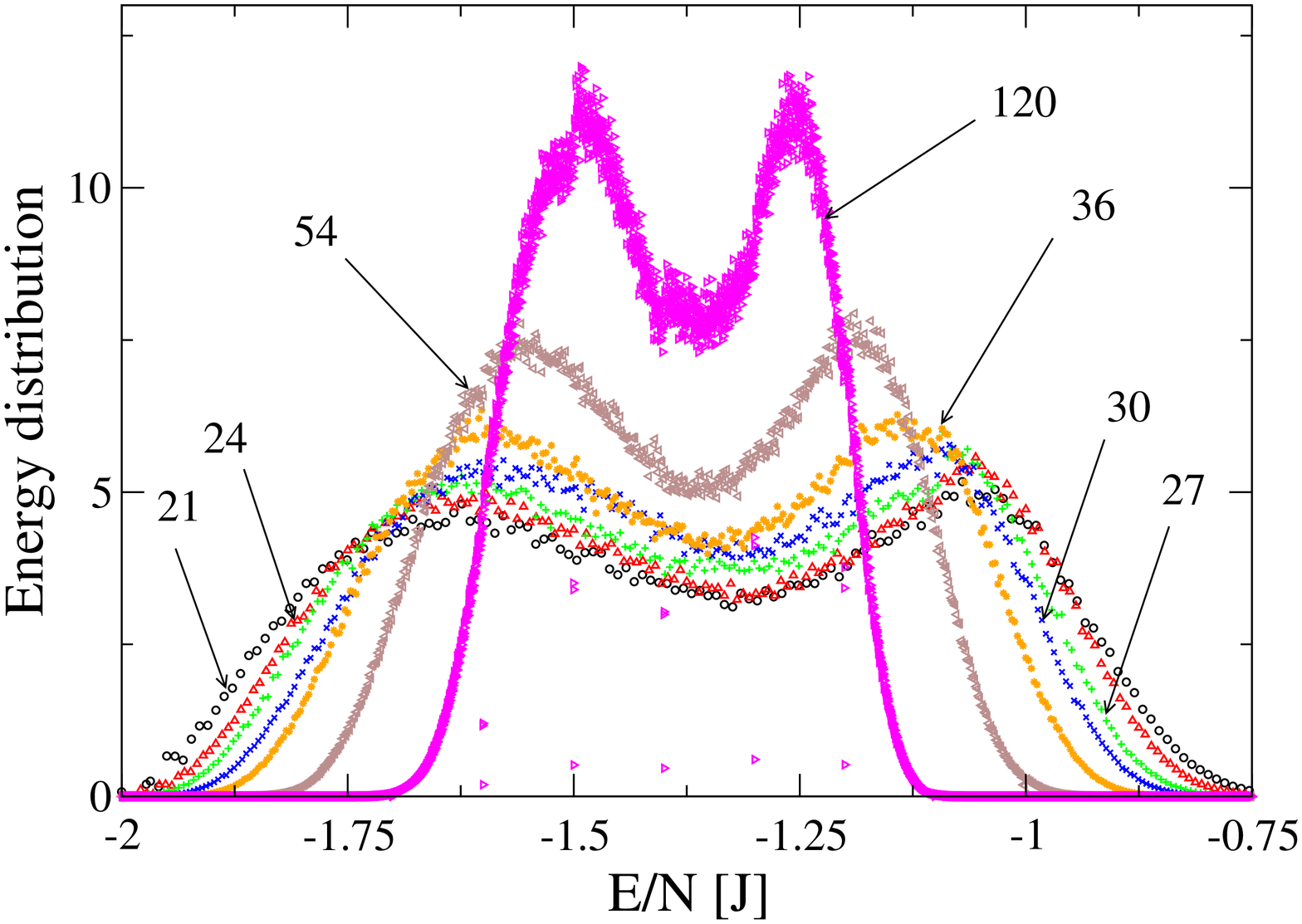}}
\caption{Critical distribution calculated at $T_{C_{\rm{max}}}$ for the pure BW model. The lattice sizes are denoted by arrows. The $L=120$ data suffers from the
systematic errors resulting from the DOS calculations for large systems.}
\label{fig:H}
\includegraphics[height=9.0cm,angle=-90]{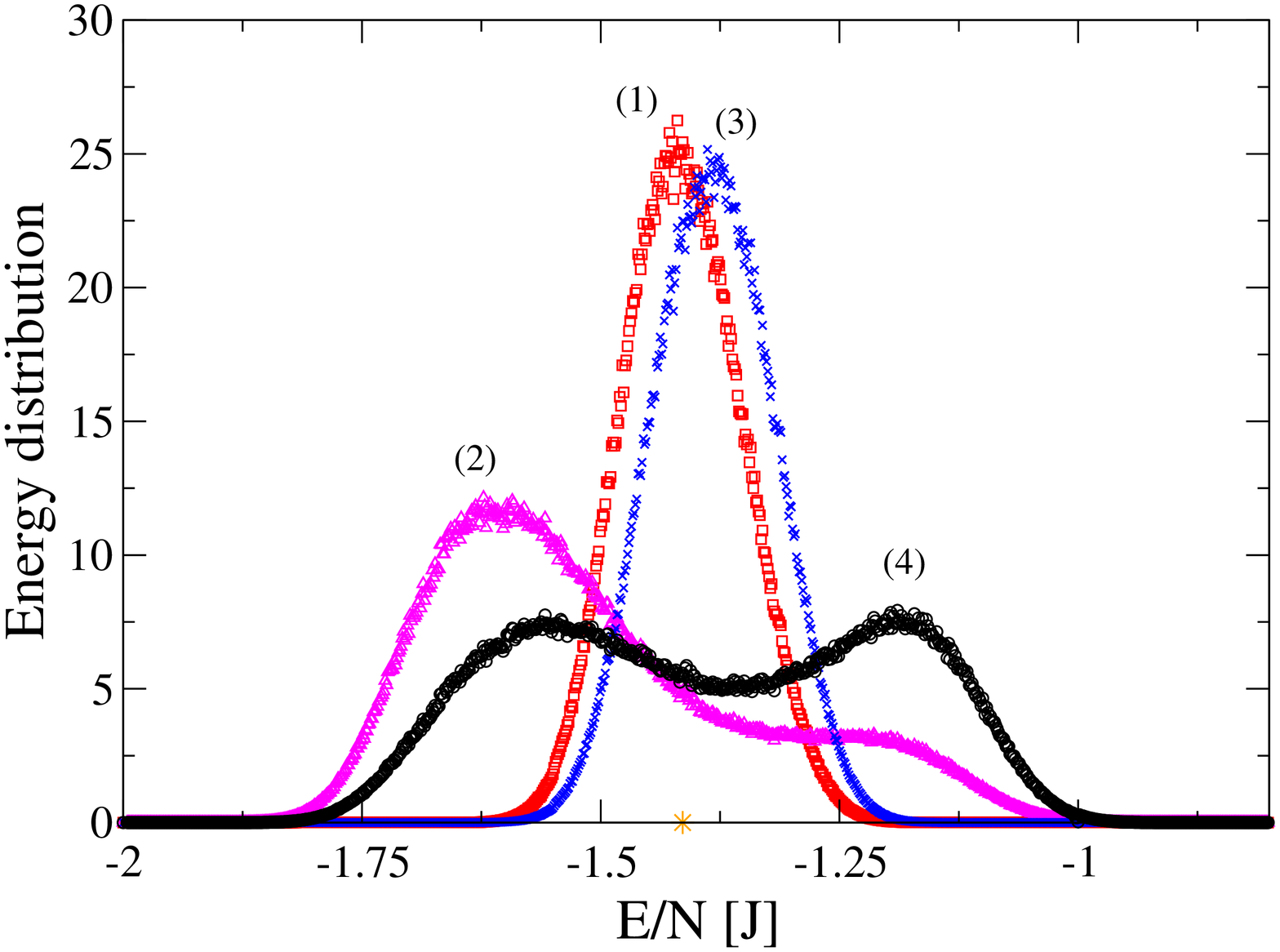}
 \caption{Energy distributions at $T_{C_{\rm{max}}}^{\rm{Ising}}=2.28948 J/k_B$, 
at $T_{C_{\rm{max}}}^{\rm{BW}}=2.27549 J/k_B$ and at the exact transition point $T_c$ on the same lattice with $L=54$. The numbers in parenthesis
denote: (1) Ising at $T_{C_{\rm{max}}}^{\rm{BW}}$, (2) BW at $T_c$,
(3) Ising at $T_{C_{\rm{max}}}^{\rm{Ising}}$, (4) BW at $T_{C_{\rm{max}}}^{\rm{BW}}$.
Note the distribution at the exact transition point (2) with the ratio of $r\simeq4$ between the pronounced peak on the left and the "hump" on the right~\cite{histogram2}.
The asterisk denotes their common critical energy $U_c=-\sqrt2J$.}
\label{fig:dos}
\end{center}
\end{figure}
The positions (energies) of the peaks are found to scale
with $L^{-1}$~\cite{Kosterlitz} as seen in Fig.~\ref{fig:E1}, 
and are expected to eventually intersect for a large enough sample.
\begin{figure}[h]
\begin{center}
\includegraphics[height=9.00cm,angle=-90]{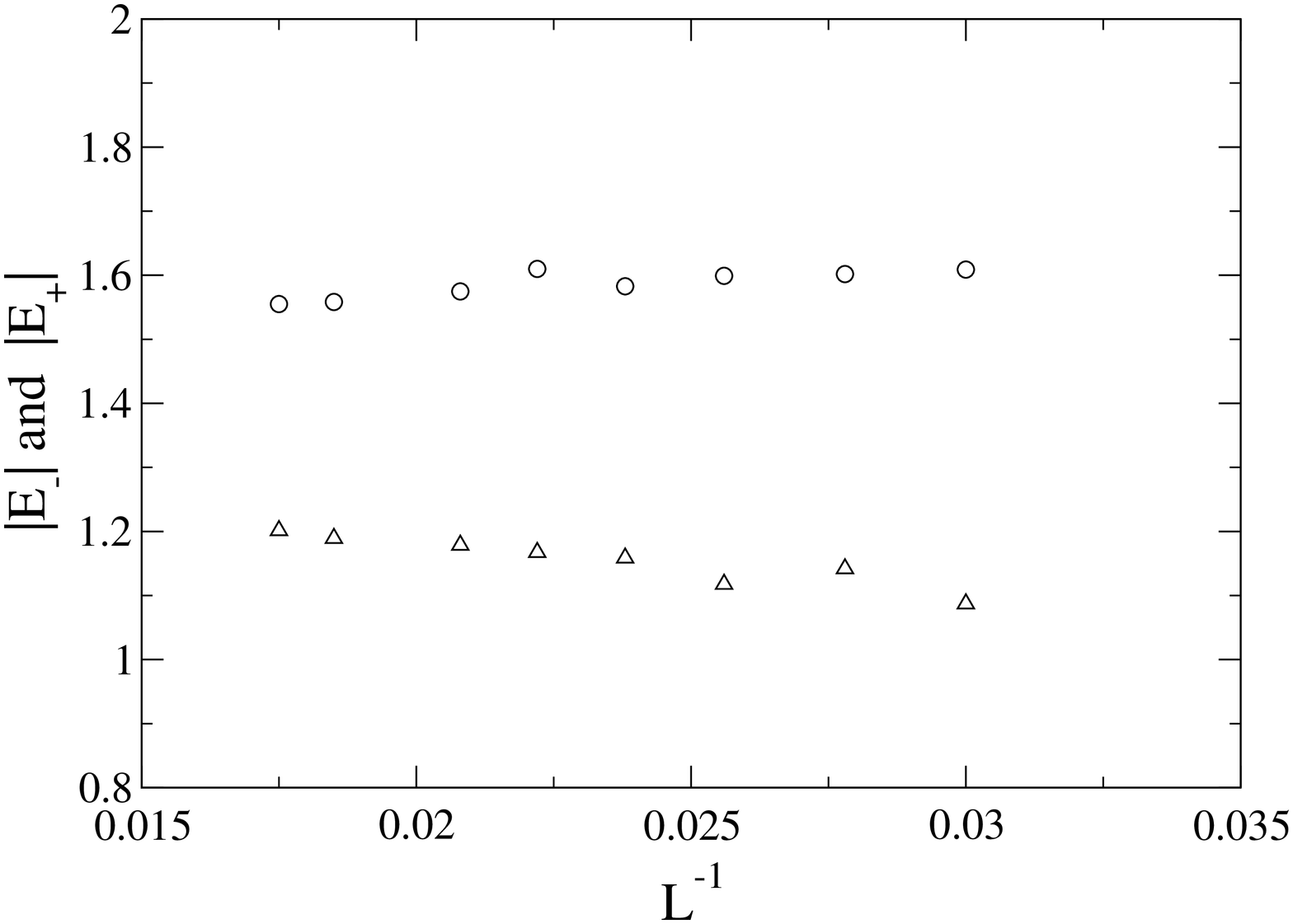}
\caption{Variation of the energy distribution's two maxima positions with
$L^{-1}$. The ``disordered'' energies are denoted by ($\triangle$) and the
``ordered'' energies by ($\circ$).}
\label{fig:E1}
\end{center}
\end{figure}

A comparison between the DOS of the Ising model and the BW model (Fig.~\ref{fig:DOS}) shows a significant difference between the two models.
Although they have approximately the same number of different energy levels ($N-1$ for Ising and $N-3$ for BW), the function $\ln g$, appears to be concave everywhere on the interval $[-2,0]$ for the Ising model, while, for the BW graph this may not be so. This suggests an explanation for the appearance of the two peaks
which is also consistent with the fact that they have the same height:
The condition that the distribution will have $extrema$ is satisfied by
\begin{equation}
 d(\ln g)/dE=1/kT.
\label{eq:Tc}
\end{equation}
If, at  $T_{C_{\rm max}}$,
 Eq.~(\ref{eq:Tc}) has locally,
a solution $f_1(E)=E/kT_{C_{\rm max}}+C_1$ tangent to $\ln g$ at $E_-$ and $E_+$,
and another solution $f_1(E)=E/kT_{C_{\rm max}}+C_2$ tangent to $\ln g$ at $U_c(L)$ (at the shifted
critical energy, or the minimum between the peaks), then the distribution will have two peaks with
equal height satisfying 
\begin{equation}
p(E_-)=p(E_+)=e^{C_1},
\end{equation}
as seen in Fig.~\ref{fig:H}.
This is essentially a finite size effect and 
should be recovered by a large enough sample, to give an "Ising like" concave everywhere DOS function, and a single peaked distribution as its consequence.

We further calculated the specific heat for each lattice size and then plotted its maximal value $C_{\rm{max}}$ versus $L$.
We see in Fig.~\ref{fig:C} a very nice agreement between the calculated data and the second order ansatz
$C_{\rm max}(L)\propto L^{\alpha/\nu}$, with $\alpha/\nu=1$, even for very small lattices $(L=6)$.

\begin{figure}[h]
\begin{center}
\includegraphics[height=9.0cm,angle=-90]{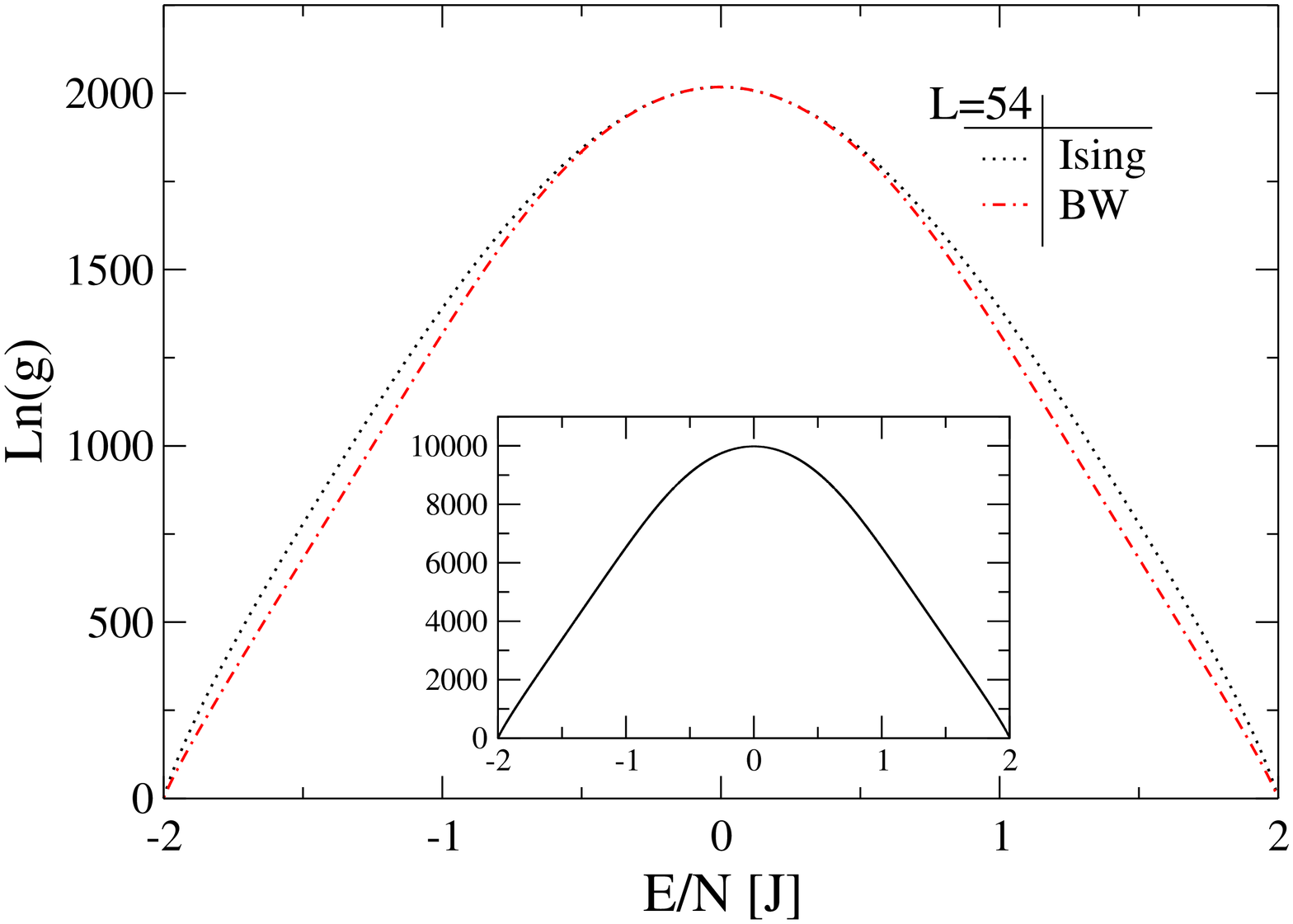}
\caption{DOS of BW and Ising models on the $L$=54 lattice. The function $\ln g$ appears to be concave "everywhere" in $[-2,0]$ for the Ising model, while this may not be so in the BW case.
A plot of $\ln g(E)$ versus $E$ for a larger BW system $(120\times120)$ is
given in the inset.}
\label{fig:DOS}
\includegraphics[height=9.0cm,angle=-90]{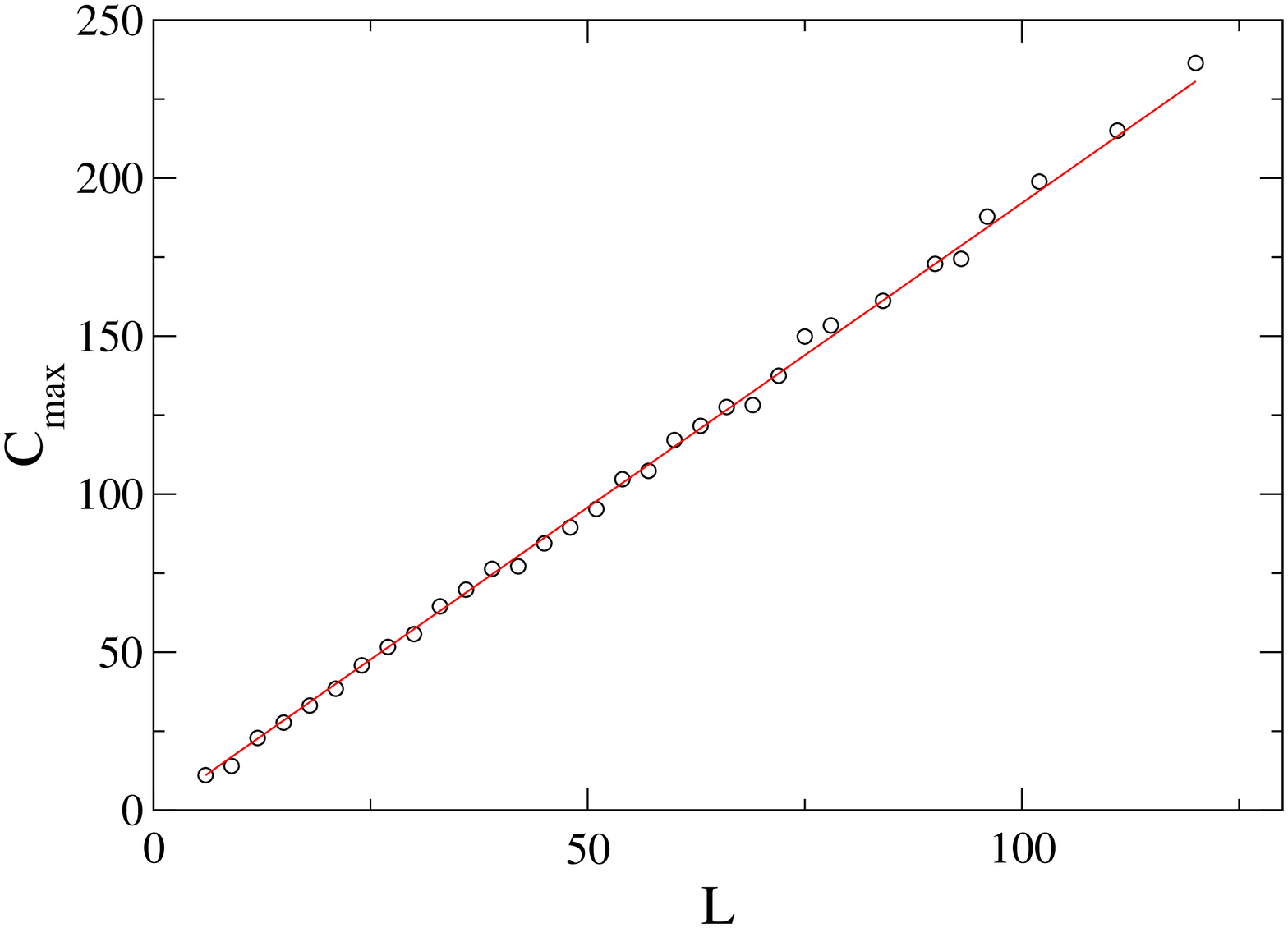}
\caption{Scaling of the specific heat maxima with the length $L$ for the pure BW model. The predicted
$C_{\rm{max}}(L)\propto L$ behavior is indicated by a solid line.}
\label{fig:C}
\end{center}
\end{figure}

\begin{figure}[h]
\begin{center}
\includegraphics[height=9.0cm,angle=-90]{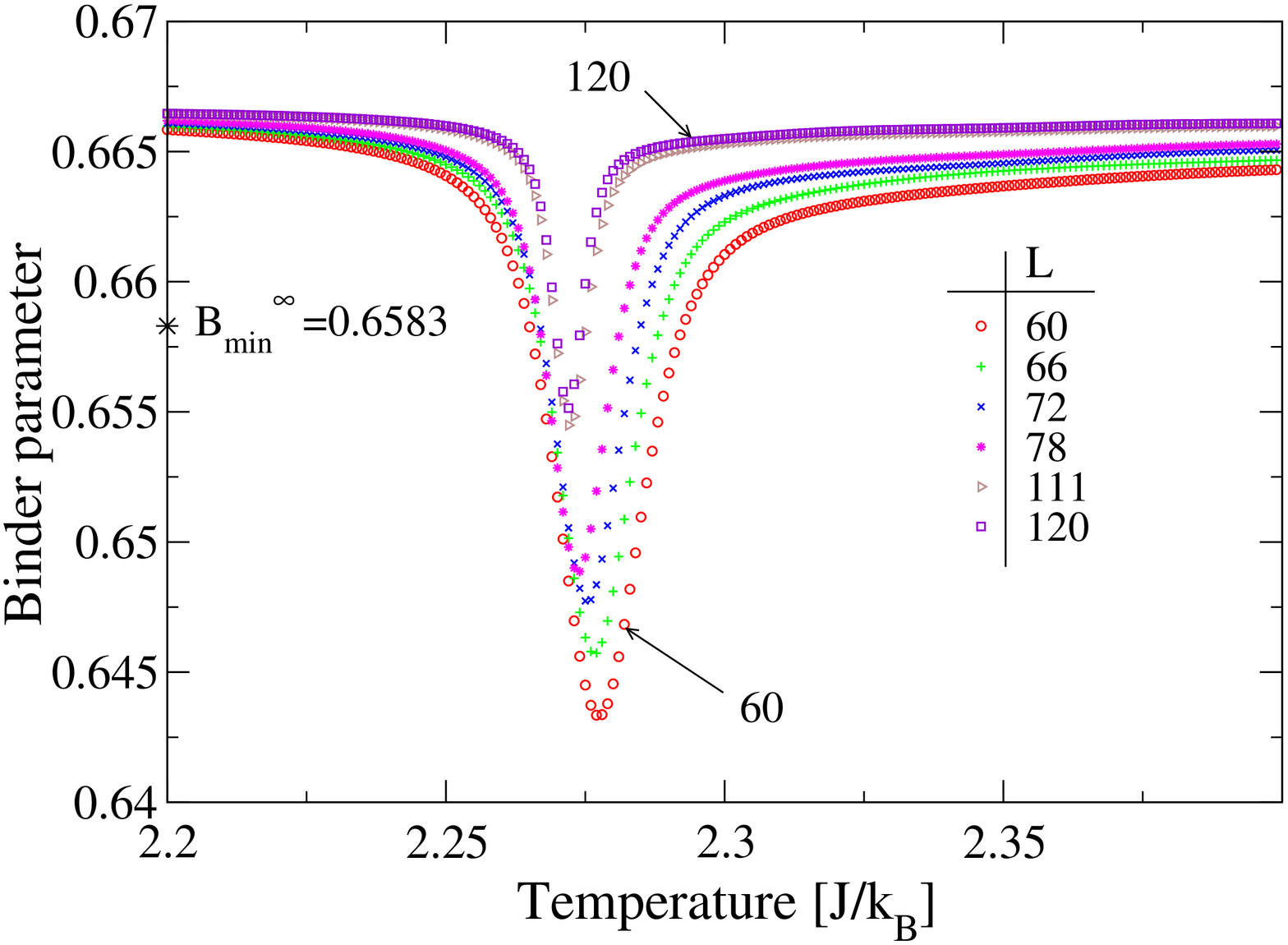}
\caption{The Binder parameter for the pure BW model versus temperature, for various lattice sizes, from top $(L=120)$ to bottom $(L=60)$ in descending order.
The Binder parameter is seen in the figure to display an inverse peak whose depth
decreases as the system size increases. 
The infinite volume upper bound $B_{\rm{min}}^\infty$ was estimated using first 
order scaling theory to $B_{\rm{min}}(L)$.}
\label{fig:B}
\includegraphics[height=9.0cm,angle=-90]{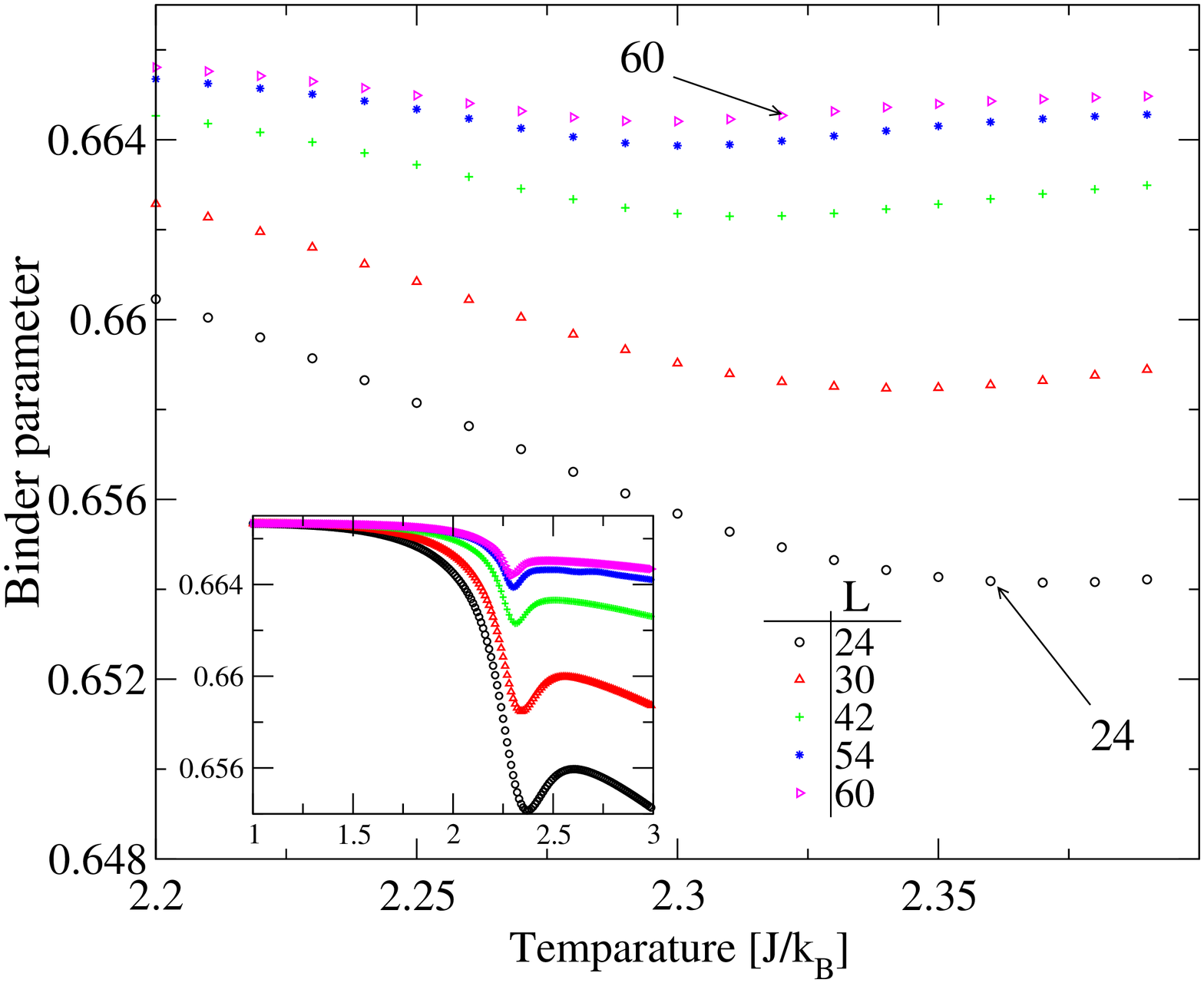}
\caption{Temperature variation of the Binder parameter for the Ising model, from top $(L=60)$ to bottom $(L=24)$ in descending order. The data in the inset is given for the
same lattices on a larger scale. The data for $L=54$ is calculated using simulated DOS. All other data is exact.}
\label{fig:B_ising}
\end{center}
\end{figure}

Another quantity of interest was the so called Binder parameter~\cite{Binder1,Binder2}
\begin{equation}
B=1-\frac{\langle E^4\rangle}{3\langle E^2\rangle^2},
\end{equation}
where $\langle...\rangle$ stands for the canonical thermal average.
When we calculated the Binder parameter,(whose plot as a function of temperature is given in Fig.~\ref{fig:B}),
we saw a sharp inverse peak that usually occurs in first order transitions~\cite{Challa,histogram2}. Another manifestation of the strong finite size effects is the very precise (though quite unreliable) estimate of
$T_c=2.2696\pm0.0004$ to the transition point we obtained, when performing first order finite
size scaling theory to the position of $B_{\rm min}$, $T_{B_{\rm min}}$. 

\begin{figure}[h]
\begin{center}
\includegraphics[height=9.0cm,angle=-90]{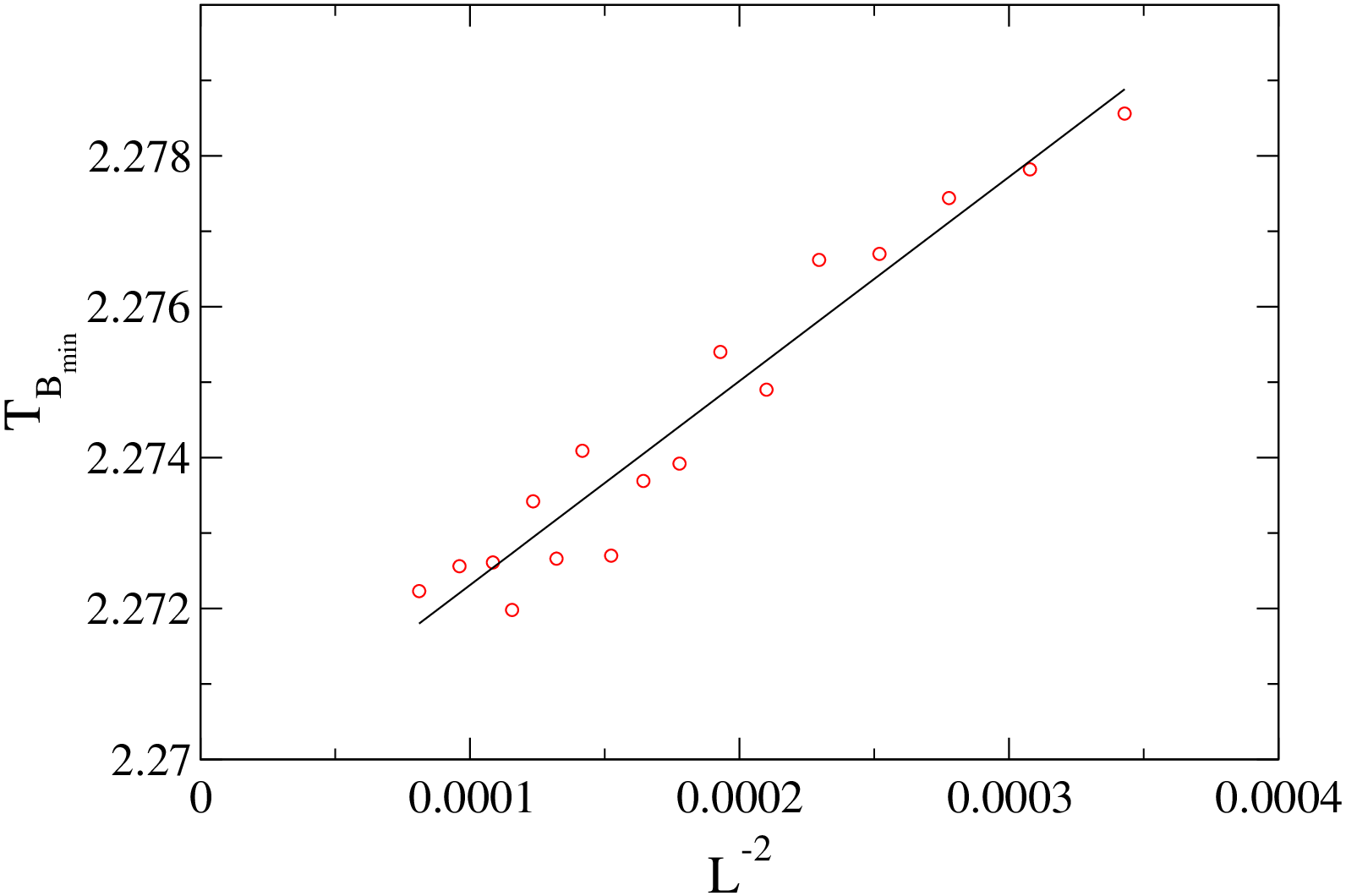}
\end{center}
\caption{Scaling of $T_{B_{\rm{min}}}$ with the inverse volume of the system for the pure BW model.}
\label{fig:Tc_Bmin}
\end{figure}

Obviously, since the transition is continuous and therefore no ordered and disordered states coexist at the transition point, the critical probability distribution in the infinite volume limit is expected to be single peaked, causing
$B_{\rm min}$ to eventually vanish with some exponent and the Binder parameter to take the trivial value
of 2/3 also at the critical point.
It was therefore convenient to repeat finite size scaling for $B_{\rm min}$
according to \begin{equation}
B_{\rm min}=\frac{2}{3}-{\cal{B}}_0L^{-\theta_B/\nu},
\label{eq:Bnir}
\end{equation}
where $\theta_B$ is an exponent yet to be determined. In Fig.~\ref{fig:Bmin_Cmax} we see the variation of the inverse distances $t_{C_{\rm max}}^{-1}\equiv\left(T_{C_{\rm max}}-T_c\right)^{-1}$ and
$t_{B_{\rm min}}^{-1}\equiv\left(T_{B_{\rm min}}-T_c\right)^{-1}$, correspond to the positions of the specific heat maxima and Binder parameter minima, respectively, with $L$. A least square fit gave a slope of $1.529\pm0.039$ for the specific heat temperature and $1.748\pm0.025$ for the Binder parameter temperature.
 In accordance with~\cite{histogram1}
\begin{eqnarray}
T_{C_{\rm max}}=T_c+A_0L^{-1/\nu}\left(1+A_1L^{-\omega_1}+\cdots\right),
\end{eqnarray}
we use the analogy
\begin{equation}
T_{B_{\rm min}}=T_c+B_0L^{-1/\nu}\left(1+B_1L^{-\theta_1}+\cdots\right) \label{eq:T_B_min},
\end{equation}
where $\omega_1$ and $\theta_1$ are correction exponents and $A_0,A_1,B_0$ and $B_1$ are amplitudes determined from simulations. It is therefore evident that
$T_{B_{\rm min}}$ displays a large correction-to-scaling term ($\theta_1\simeq0.25$),
in contrary to the resulting $1/\nu$ scaling from the $T_{C_{\rm max}}$ fit, which is in fair agreement with the exact 3/2 value, and which is also consistent with the scaling of $C_{\rm max}$. It is also evident, however, from Fig.~\ref{fig:Cmax_Bmin} and Eq.~(\ref{eq:Bnir}), that $\alpha$ and $\theta_B$ have the same value. Similar exact and simulational calculations of the Binder parameter for the
Ising model on the same temperature scale (Fig.~\ref{fig:B_ising}) showed much
broader and less deep minima at $T_{B_{\rm min}}$, suggesting that these minima vanish with an exponent $\theta_B$ larger than the BW exponent.

\begin{figure}[h]
\begin{center}
\includegraphics[height=9.0cm,angle=-90]{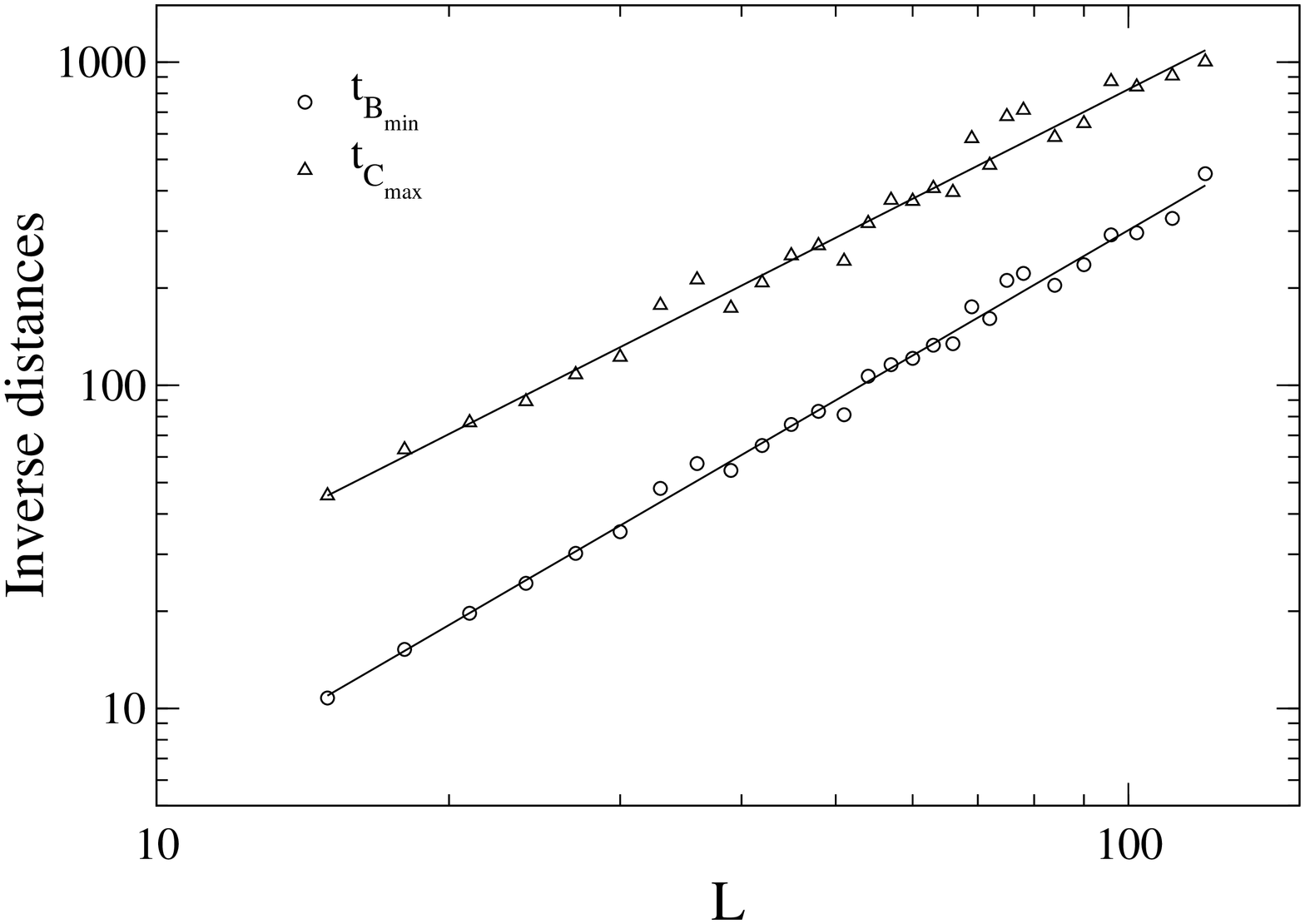}
\caption{Scaling of the inverse distances $t_{C_{\rm max}}^{-1}$ and $t_{B_{\rm min}}^{-1}$, with $L$, for the BW model. The larger slope of the Binder parameter position's fit, may be a result of the large correction term $\theta_1$
(see Eq.~\ref{eq:T_B_min}). The specific heat data was shifted for ease of reading.}
\label{fig:Bmin_Cmax}
\includegraphics[height=9.0cm,angle=-90]{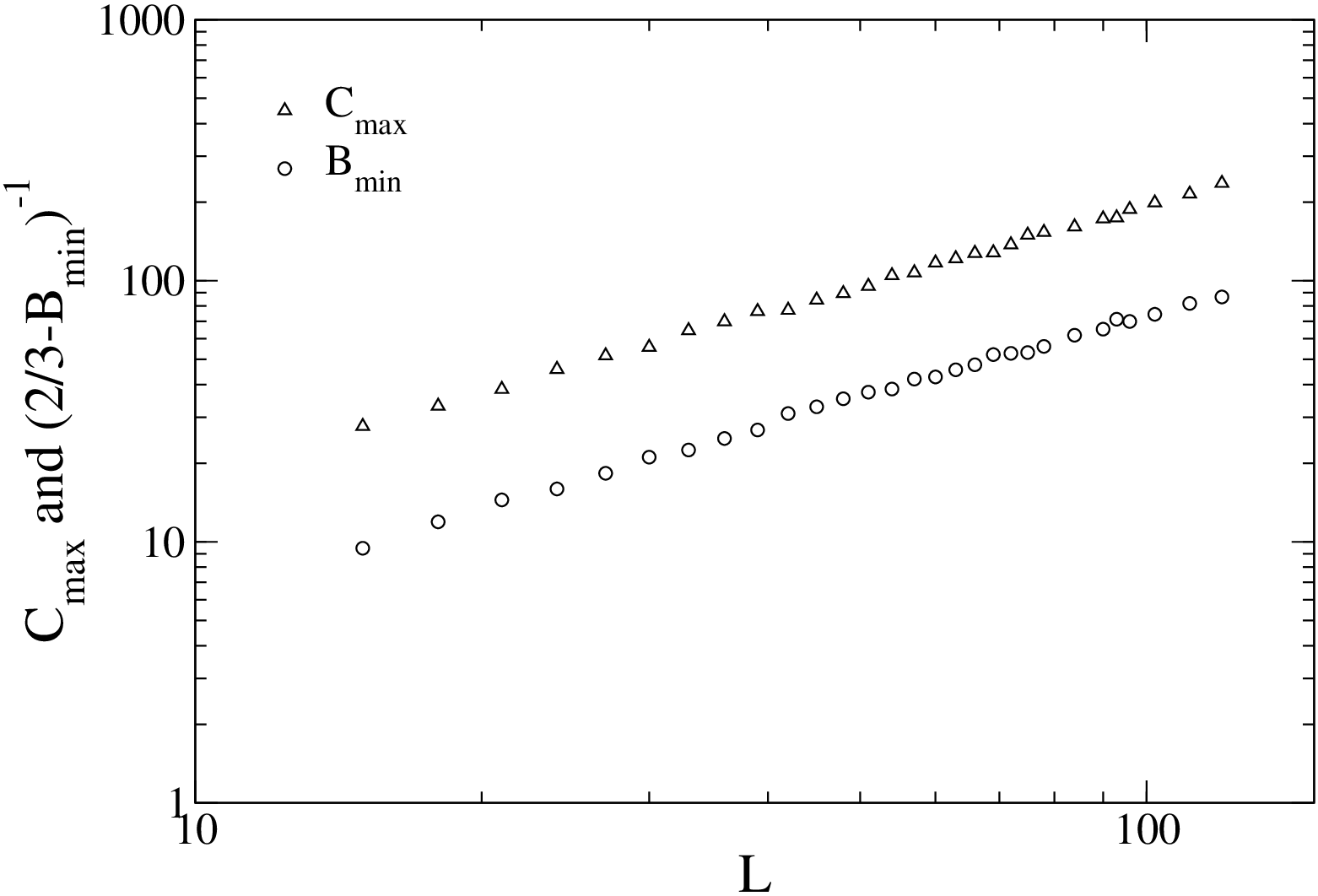}
\caption{Scaling of the quantity $\left(2/3-B_{\rm min}\right)^{-1}$ and
the specific heat maximum $C_{\rm max}$, with $L$, for the BW model.}
\label{fig:Cmax_Bmin}
\end{center}
\end{figure}

\section{The dilute BW model}{\label{sec:BW_dilute}}
Let us consider now the ferromagnetic BW model with quenched impurities. The Hamiltonian is given by
\begin{equation}
{\cal{H}}=-J\sum_{i,j,k}\epsilon_i\epsilon_j\epsilon_k\sigma_i\sigma_j\sigma_k.
\end{equation}
We studied systems with lengths $L$ between 18 and 36. We kept concentrations of $x=0.8$ for $L=18$ and of $x=0.9, 0.95$ and $x=0.97$
for $L=33$, fixed, and let them vary around $x=0.9$ for $L=33$. The data for $L=24$ was calculated for concentrations varied around
different values from $x=0.85$ to $x=0.97$.
In Fig.~\ref{fig:DOS_36_pure_dilute} we compare the DOS of the pure and dilute BW models. The apparent crossover to a manifestly clear second order transition may give rise again to a concave everywhere form of $\ln g$, already seen for the Ising model
in Fig.~\ref{fig:DOS}.
The energy levels differ now only in the amount of $2J$ and can take even or odd values for the same lattice size, depending on
the vacancy distribution.
We then performed a calculation similar to that made above for the dilute Ising model, of $T_{C_{\rm{max}}}$, to obtain
the $T_c(x)$ critical line on a lattice
with $L=24$, and then fitted the high concentration data into a continuous (dotted) line (Fig.~\ref{fig:BWdilute}).  All the data except for the $L=18$ with a vacancy concentration
of 0.2, which was, as for the dilute Ising model, unreliable because of 
relatively high dilution, fell very well on the dotted line. This may suggest that the
critical behavior is rather universal for large enough concentrations, because due
to the special connectivity of the BW model, one would expect smaller energy fluctuations
and therefore a larger scatter of data for large enough vacancy concentrations, whilst the dilute BW data seems to agree with the
dilute Ising data for concentrations of $x\geq0.9$. Of course, in order to make
definitive statements about universality, larger samples would be needed than 
those used here.
\begin{figure}
\begin{center}
\includegraphics[height=9.0cm,angle=-90]{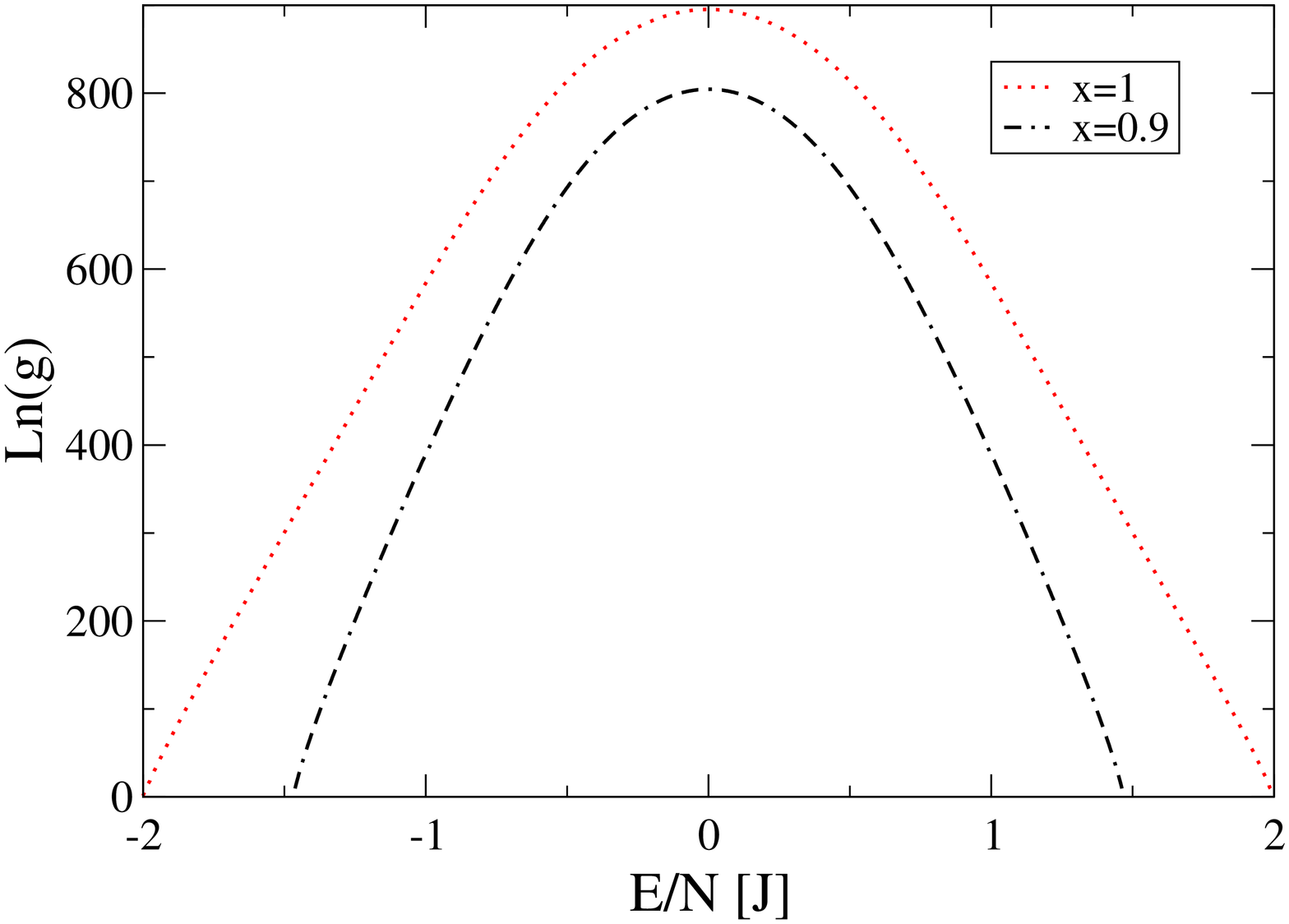}
\caption
{DOS for a pure (upper curve) and for a dilute (lower curve) BW model with $x=0.9$, on an $L=36$ lattice.}
\label{fig:DOS_36_pure_dilute}
\end{center}
\end{figure}
\begin{figure}[h]
\begin{center}
\includegraphics[height=9.0cm,angle=-90]{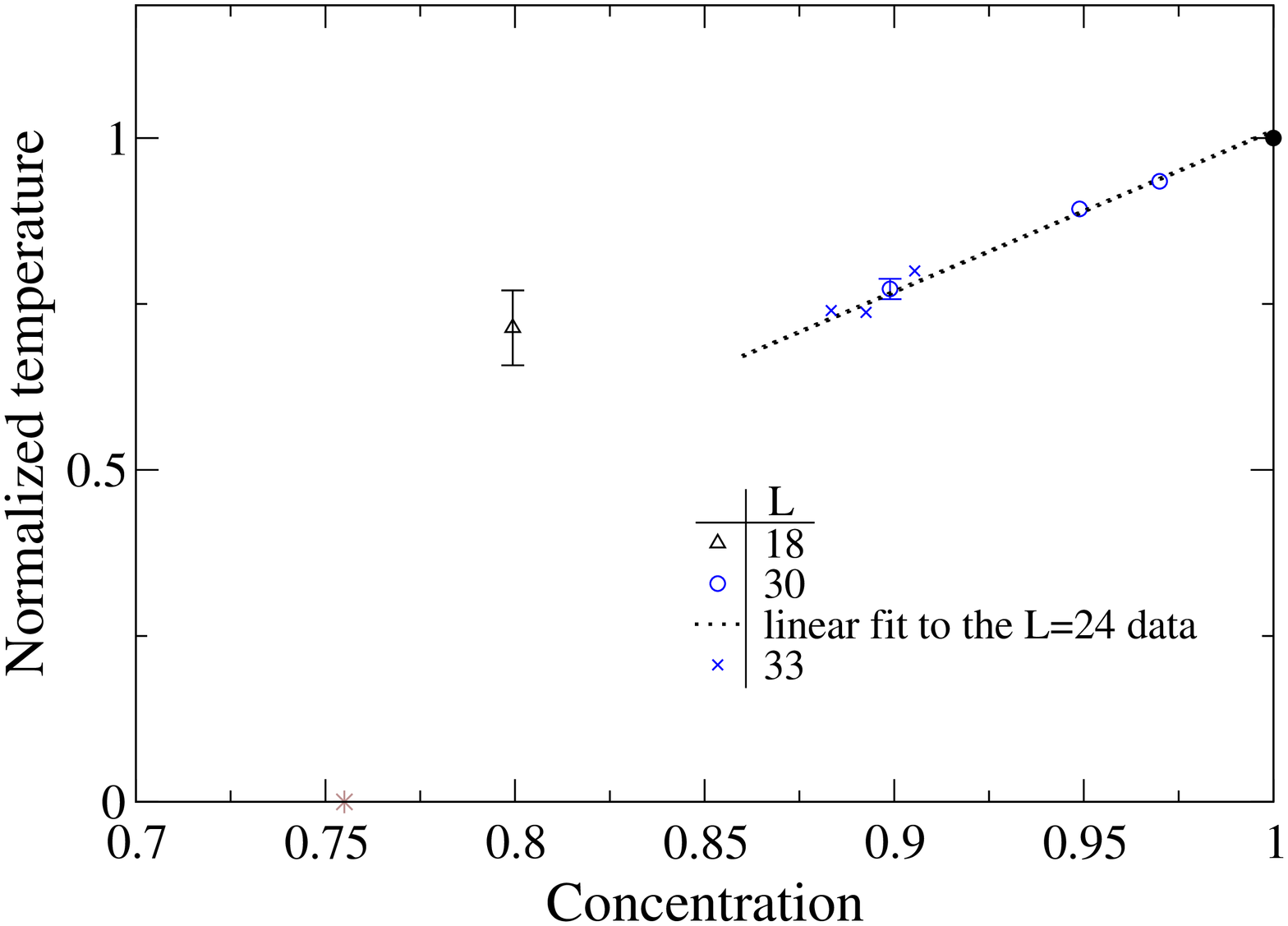}
\caption{Normalized critical temperature for fixed concentrations
on different lattice sizes for the dilute BW model.}
\label{fig:BWdilute}
\end{center}
\end{figure}

We performed a rough finite size scaling for the specific heat maxima at a concentration of $x=0.9$,
using the three points measured for $L=33$ that were averaged and the other data collected for
fixed concentrations.
Novotny and Landau~\cite{Novotny} predicted $\alpha/\nu\simeq0$ for a concentration of 0.9.
Our results, presented in Fig.~\ref{fig:nu=0} also indicate, at least qualitatively, a significant change in $\alpha$.
Since spatial correlations become smaller and hence $\nu$ becomes smaller,  the value of $\alpha$ substantially decreases, thus indicating an "Ising like" singularity at the finite lattice transition point. Moreover, the Harris criterion for the diluted case is hereby confirmed.
\begin{figure}
\begin{center}
\includegraphics[height=9.0cm,angle=-90]{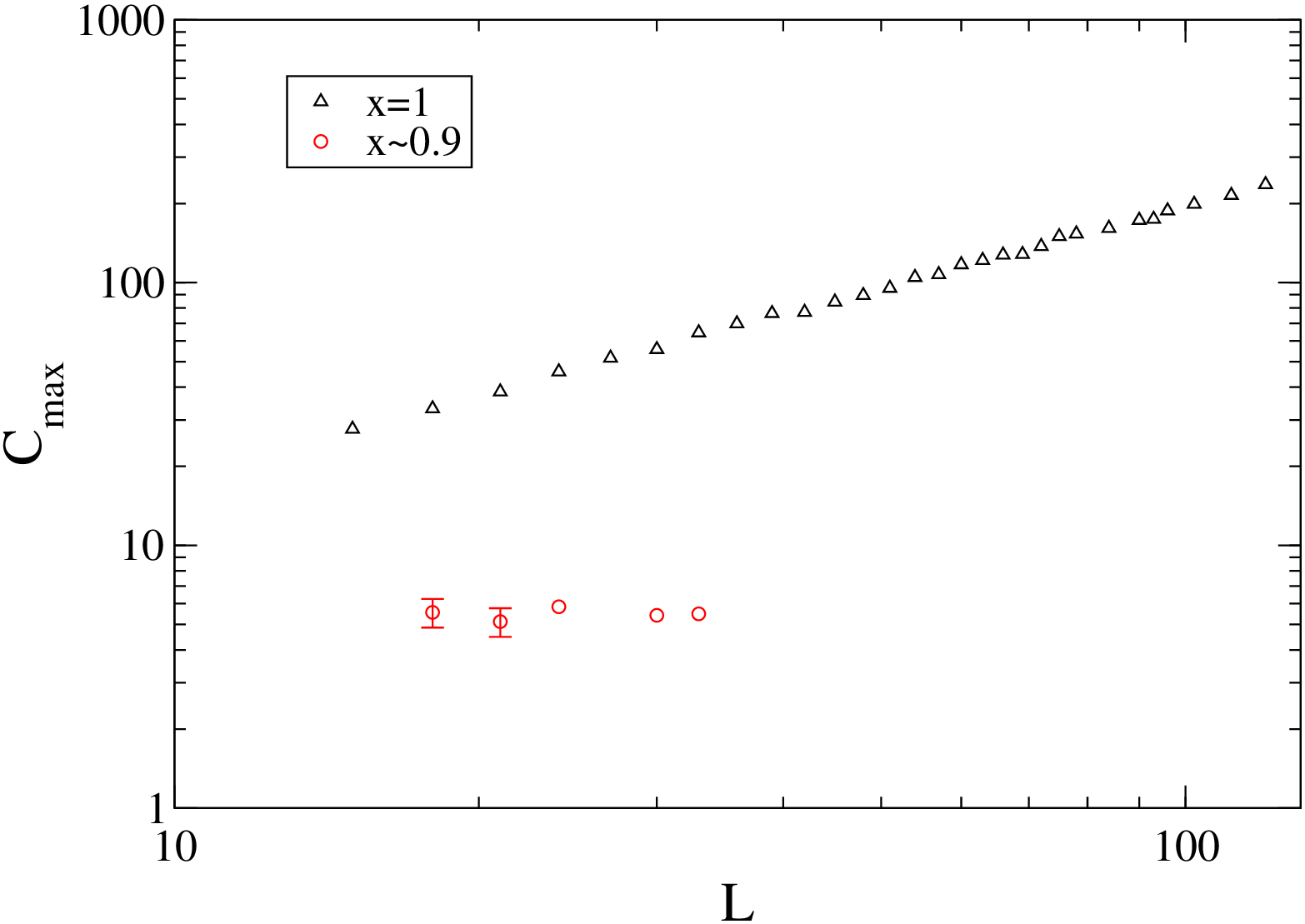}
\caption{Finite size scaling of $C_{\rm max}$ with $L$ for the pure and dilute BW model. The data for the dilute model
reveals an $\alpha$ exponent close to zero.}\label{fig:nu=0}
\end{center}
\end{figure}
Another question of interest was the influence of vacancies on the nature of the transition.
In order to make a statement regarding this question we plotted in Fig.~\ref{fig:pBWdilute} the energy distribution
for different concentrations. We see clearly and unsurprisingly that lowering the concentration causes the doubly peaked distribution
to vanish and become a singled peaked one with a narrower width centered away from $U_c$. It may then be plausible to say that
in contrast to energy fluctuations which become negligible at sufficiently low concentrations,
magnetic fluctuations increase with increasing dilution and the transition is manifestly
second order.
\begin{figure}
\begin{center}
\includegraphics[height=9.0cm,angle=-90]{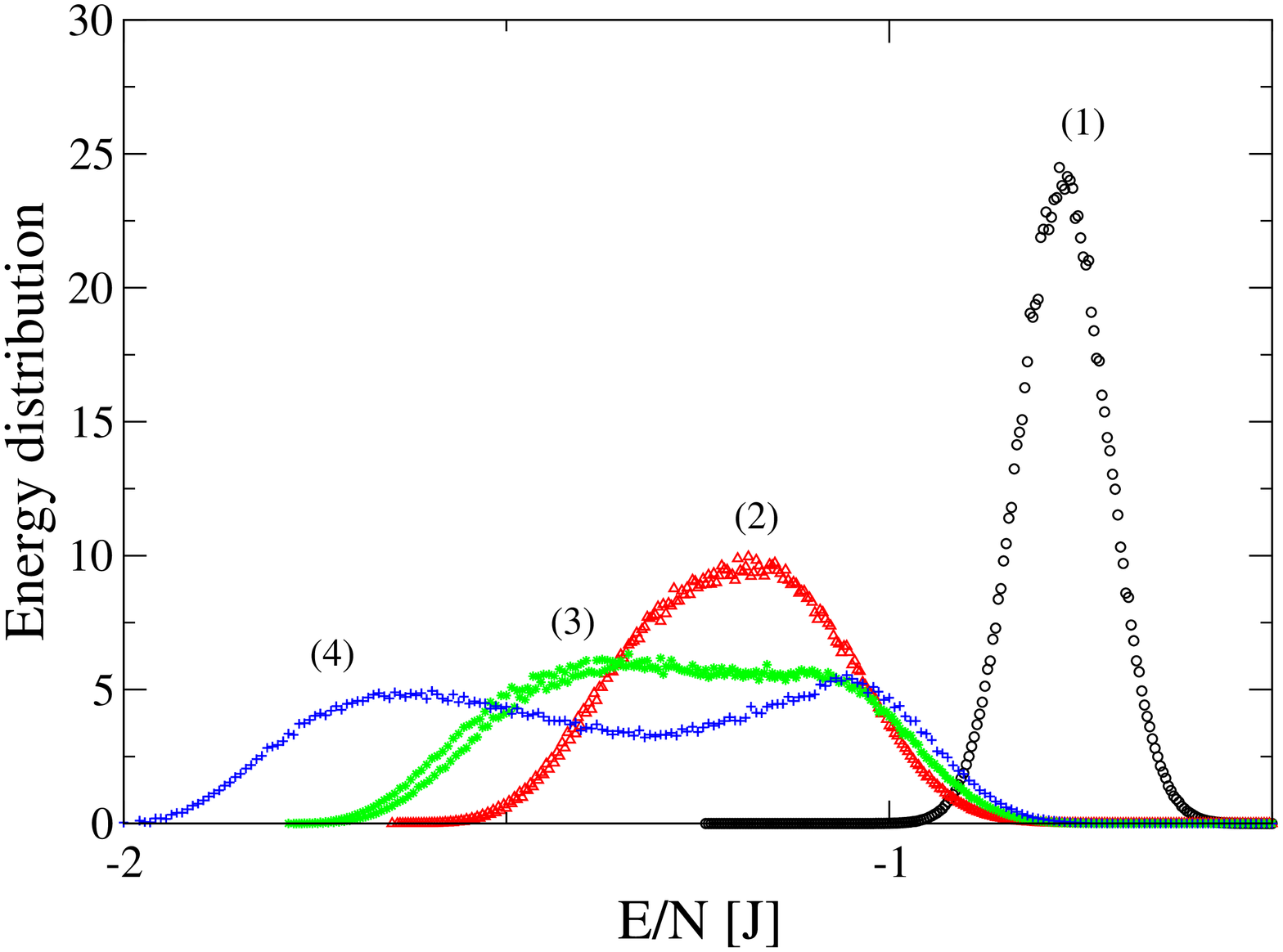}
\caption{Critical energy distribution for various concentrations and critical temperatures, calculated on a $L=24$ lattice. 
The numbers in parenthesis denote: (1) $x=0.85$; $T_c=1.74926$, (2)  $x=0.95$; $T_c=1.93379$,
(3)  $x=0.97$; $T_c=2.07671$, and (4) $x=1$ (pure); $T_c=2.29164$.
 The distribution is seen to become sharper and narrower when the concentration is reduced.}
\label{fig:pBWdilute}
\end{center}
\end{figure}
\section{Conclusions}{\label{sec:discussion}}
Our simulations have shown that the WL sampling is a very accurate algorithm.
The thermodynamic quantities resulting from the calculated $g(E)$, which yield reasonable quality critical data, provide good evidence for this.

Our results show that the pure Baxter-Wu model is strongly influenced by finite size effects and corrections to
scaling. The scaling of the specific heat maxima is in excellent agreement with the second order form $C_{\rm max}\propto L^{\alpha/\nu}$, even for small lattices, and no correction terms are observed. The Binder parameter, however, displays large
minima for small samples, thus incorrectly could be thought of as a "first order" scaling field. 
It is an "irrelevant" field in the sense that it gives no additional information about the universal exponent
$\nu$, but rather vanishes with an exponent $\theta_B$.
This exponent is also evident in the Ising model and is presumably larger for this model. The vanishing inverse peak in both models states that the energy distribution approaches a delta function in the thermodynamic limit, although it is essentially non-Gaussian.
The doubly peaked shape of the latter is rather peculiar. One would usually expect a single peaked distribution which 
becomes narrower, the closer to criticality one is. This shape is essentially a finite size effect due to the large fluctuations between the ferromagnetic and ferrimagnetic clusters formed in the vicinity of the transition
point, and will eventually vanish in the thermodynamic limit. 
The WL method is also very successful when applied for the dilute BW model even for small lattices, 
both in terms of the critical isotherm in temperature-concentration plane for a weak dilution, and probability distribution.
A crossover to a single peaked critical distribution is clearly seen when decreasing the concentration of spins,
and a single peaked distribution is evident at a concentration of $x=0.85$. This is a result of the formation of isolated
domains causing relatively small energy fluctuations around the critical energy. 

It would be interesting in the future to use larger lattices to confirm our explanations of the finite 
size problems, The relatively high accuracy of the WL method for small dilute systems could be applied in the future to study 
disorder in other models.

\section*{Acknowledgments}
We thank Prof. D. Stauffer, Prof. D.P. Landau, Prof. M.E. Fisher and Prof. W. Janke for useful comments and suggestions.
We would like to thank E. Warszawski and I. Klich for helpful discussions.
We thank the BSF for generous support 
throughout this project. The financial support of the Technion is also 
gratefully acknowledged. \\ 

\end{document}